\documentclass[cits]{PoS}
\pdfoutput=1

\usepackage{microtype}
\usepackage{amsmath}

\newcommand{\bk}{(\beta,\,\kappa)}
\newcommand{\bkwk}{(1.60,\,0.1371)}
\newcommand{\bkst}{(1.58,\,0.1385)}

\DeclareMathOperator{\Prob}{Prob}

\DeclareMathOperator{\re}{Re}
\DeclareMathOperator{\im}{Im}
\DeclareMathOperator{\dd}{d\!}

\title{Zeros of QCD partition function from finite density lattices}

\ShortTitle{Zeros of QCD partition function from finite density lattices}

\author{
  \speaker{Xiao-Yong Jin}$^{\dag a}$, 
  Yoshinobu Kuramashi$^{abc}$,
  Yoshifumi Nakamura$^a$,
  Shinji Takeda$^{ad}$,
  and
  Akira Ukawa$^c$
  \vspace{1em}\\
  \llap{$^a$}RIKEN Advanced Institute for Computational Science,\\
  Kobe, Hyogo 650-0047, Japan
  \vspace{1em}\\
  \llap{$^b$}Graduate School of Pure and Applied Sciences,\\
  University of Tsukuba,\\
  Tsukuba, Ibaraki 305-8571, Japan
  \vspace{1em}\\
  \llap{$^c$}Center for Computational Sciences,\\
  University of Tsukuba,\\
  Tsukuba, Ibaraki 305-8577, Japan
  \vspace{1em}\\
  \llap{$^d$}Institute of Physics,\\
  Kanazawa University,\\
  Kanazawa 920-1192, Japan
  \vspace{1em}\\
  \llap{$^\dag$}Email: \email{xjin@riken.jp}
}

\abstract{

  Partition function zeros steer the critical behavior of a system.
  Studying four-flavor lattice QCD at finite temperature and density
  with the Wilson-clover fermion action and the Iwasaki gauge action
  using a phase-quenched fermion determinant, we combine statistics
  from multiple chemical potentials to improve sampling of the
  configuration space, and aim at unraveling the movement of zeros in
  finite systems.  Preparing for further investigations, we discuss
  methods and criteria used to sieve through complex parameter space
  spanned by $(\re\mu,\, \im\mu)$ and $(\re\mu,\, \im\beta)$, and
  present statistically robust zeros of the partition function.

}

\FullConference{31st International Symposium on Lattice Field Theory LATTICE 2013\\
  July 29 -- August 3, 2013\\
  Mainz, Germany}

\begin{document}

\section{\label{sec:introduction}Introduction}

Analyses of partition function zeros may pave the way to the critical
point of QCD at finite temperature and
density~\cite{Fodor:2001pe,Fodor:2004nz}---if executed with
care~\cite{Ejiri:2005ts}.  Danger lurks in the thermodynamic limit
with finite statistics, where zeros would appear on the real axis, a
manifestation of the ``sign problem''.  With sufficient statistics,
however, we can investigate unimpaired zeros of finite systems and
infer the thermodynamic limit.

We study the finite temperature and density ensembles generated with a
phase-quenched fermion determinant~\cite{Jin:2013wta}, using four
fermion flavors of the Wilson-clover action and the Iwasaki gauge
action.  Reweighting from simulations at multiple $\mu$
values,\kern-2pt\footnote{We omit the lattice constant, $a$, for
  brevity.} we locate zeros of the partition function.  We present
methods and criteria for a statistically robust zero from combined
simulations, and discuss our results.

\section{\label{sec:ensembl-rewe}Ensembles and reweighting}
  
We use multi-ensemble reweighting~\cite{Ferrenberg:1989ui} to combine
ensembles with various chemical potentials, $\mu$, at two combinations
of the bare lattice coupling and the inverse quark mass, $\bk = \bkwk$
and $\bkst$, with lattices of a temporal length, $N_t=4$, and spatial
volumes, $V = 6^3$, $6\cdot6\cdot8$, $6\cdot8\cdot8$, $8^3$, and
$10^3$ (only at the stronger coupling).  To maximize the overlap and
reduce the reweighting noise, we employ ensembles with $\mu$ values
close to the transition/cross-over region, listed in
Table~\ref{tab:configs}.  We compute action differences due to
changing $\mu$ using the first four terms from the Taylor expansion of
the logarithm of the fermion determinant with respect to $\mu/T$.
Allton et al.~\cite{Allton:2005gk} implemented $\mu$-reweighting from
zero density ensembles, whereas we reweight from multiple ensembles
simulated at values of $\mu$ that are close to the region of interest.

\begin{table}[b]
  \centering
  \begin{tabular}{lrrrp{6mm}lrrrr}
    \cline{1-4}\cline{6-10}
    $V$             & $\mu=0.2$ & $0.205$ & $0.21$ &  & $V$             & $\mu=0.13$ & $0.14$ & $0.15$ & $0.16$ \\
    \cline{1-4}\cline{6-10}
    $6^3$           & 16000     & 16000   & 16000  &  & $6^3$           & 5000       & 5000   & 5000   & 5000   \\
    $6\cdot6\cdot8$ &           & 32000   &        &  & $6\cdot6\cdot8$ & 5000       & 5000   & 5000   & 5000   \\
    $6\cdot8\cdot8$ &           & 90000   &        &  & $6\cdot8\cdot8$ & 5000       & 13000  & 13000  &        \\
    $8^3$           & 90000     & 120000  & 90000  &  & $8^3$           & 27500      & 27500  & 27500  &        \\
                    &           &         &        &  & $10^3$          &            & 34780  & 34280  & 11390  \\
    \cline{1-4}\cline{6-10}
  \end{tabular}
  \caption{\label{tab:configs}Number of configurations used
    in multi-ensemble reweighting for $\bk = \bkwk$ on the left and
    $\bk = \bkst$ on the right.}
\end{table}

\begin{figure}
  \centering
  \includegraphics[scale=0.93]{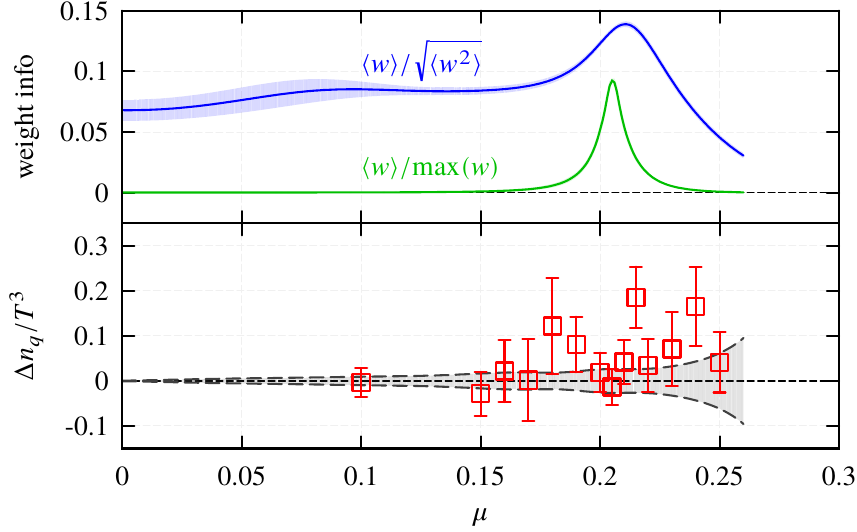}
  \caption{\label{fig:quark-numd-reweight}The lower panel shows, for
    $\bk = \bkwk$ with $V=8^3$, the quark number density from each
    ensemble subtracted by the multi-ensemble ($\mu=0.2$, $0.205$, and
    $0.21$) reweighted value, with the reweighted standard deviation
    shown as the gray band.  The upper panel shows the weight
    statistics, with the partially indiscernible shadow band
    representing the corresponding standard deviation, where $w$
    represents the real part of the weight and $\langle\cdot\rangle$
    indicates multi-ensemble average.}
\end{figure}

Reweighting results from these ensembles near transition/cross-over
region reproduce phase-reweighting only results from ensembles
simulated at other $\mu$ values.  Shown in
Figure~\ref{fig:quark-numd-reweight}, at each simulation point with
$\bk = \bkwk$ and $V=8^3$, the phase-reweighted quark number density
deviates within the statistical uncertainty from the multi-ensemble
$\mu$-reweighted value.

\section{\label{sec:part-funct-zeros}Partition function zeros and
  reliability}

For a finite system simulated with real parameters, zeros of the
partition function manifest themselves in the vanishing of both real
and imaginary parts of the average weight from reweighting with
complex parameters.  In practice, by reweighting with a real
parameter, $p$, and an imaginary part of some parameter, $q$, we
minimize the squared absolute value of a ratio of total weights, $w$,
\begin{equation}
  \label{eq:2}
  \mathcal{Z}_{\text{norm}} =
  \frac{\sum_U w(p,\, iq;\, U)}{\sum_U w(p,\, iq=0;\, U)}\,,
  \quad
  \text{where $\textstyle\sum_U$ is over all configurations.}
\end{equation}
It is effectively a ratio of partition functions with excessive
oscillating terms removed.

In this paper, we report on zeros of partition functions in the
parameter space of $(p,\, q)$ being either $(\re\mu,\, \im\mu)$ or
$(\re\mu,\, \im\beta)$.

\begin{figure}
  \centering
  \includegraphics[width=0.48\textwidth]{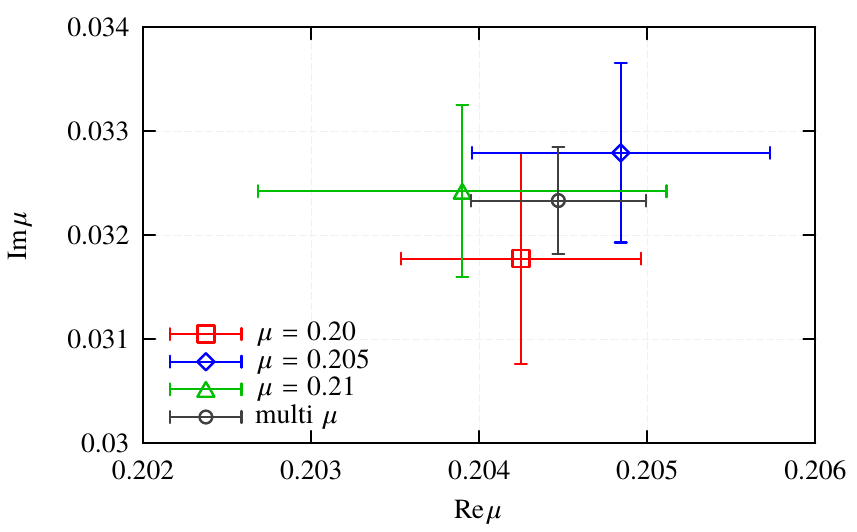}
  \hfill
  \includegraphics[width=0.48\textwidth]{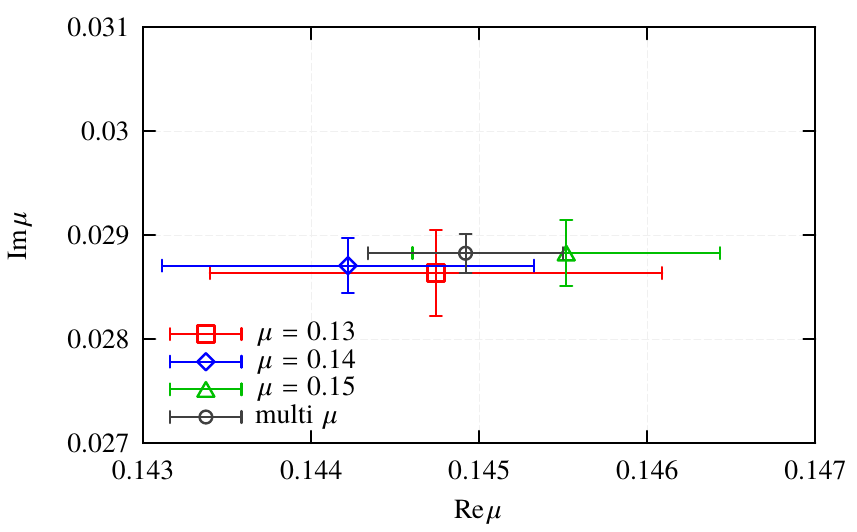}
  \caption{\label{fig:1385_mu_imu}Location of the first partition
    function zero with $V=8^3$.  Shown in the left panel is $\bk =
    \bkwk$; the right panel $\bk = \bkst$.}
\end{figure}

We observe consistency across results from reweighting of a single
ensemble and multiple ensembles.  Figure~\ref{fig:1385_mu_imu}
compares multi-ensemble reweighting to single-ensemble reweighting in
the complex $\mu$ plane.  Combining three ensembles achieves more than
a $40\%$ reduction in statistical uncertainties.

To assess the numerical reliability of the partition function zeros,
we seek constraint from the uncertainty of the partition function with
a complex parameter, $x$, reweighted from $x_0$,
\begin{equation}
  \label{eq:3}
  \mathcal{Z}'
  \approx
  \int \dd O e^{O(x-x_0)} \Prob_0[O]
  =
  \frac{1}{\mathcal{Z}_0}
  \int \dd U e^{-S_0 + O_0(x-x_0)}\,,
\end{equation}
where the observable $O=-\frac{\partial S}{\partial x}$, with
$\Prob_0[O]$ the distribution of $O$ at $x_0$.  It approximates the
partition function up to the first derivative of the action, though
the relation using $\beta$ is exact~\cite{Alves:1991qp}.  Away from
real zeros generated by two approximately Gaussian peaks, with $L$
independent and identically distributed configurations, we derive the
confidence region, $|x-x_0|$, satisfying $|\mathcal{Z}'| >
k\sigma(|\mathcal{Z}'|)$,
\begin{equation}
  \label{eq:1}
  \sum_j \left|x_j-x_{j,0}\right|^2\sigma^2(O_j)
  <
  \ln\Big[\Big(\frac{L}{k^2}+1\Big)\big/2\Big]\,,
\end{equation}
where $\sigma^2(O_j)$ is the variance of a single peak of its
distribution, with $x_j$ denoting multiple parameters.

\begin{figure}
  \centering
  \includegraphics{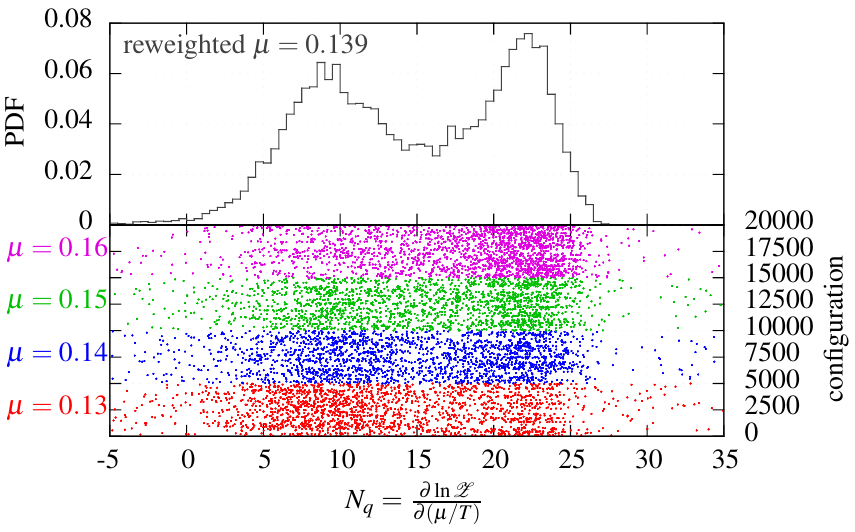}
  \caption{\label{fig:b158_nl6nt4_k01385_dist_qnumd}Distribution of
    quark number, $N_q=\frac{\partial \ln\mathcal{Z}}{\partial
      (\mu/T)}$, for $\bk = \bkst$ with $V=6^3$.  Top panel shows the
    histogram at $\mu=0.139$ by reweighting from all configurations
    shown in the bottom.}
\end{figure}

For $\mu$-reweighting, we look at the reweighted probability
distribution function of the quark number,
$N_q=\frac{\partial\ln\mathcal{Z}}{\partial(\mu/T)}$, and estimate its
width.  Figure~\ref{fig:b158_nl6nt4_k01385_dist_qnumd} shows an
example for $\bk = \bkst$ with $V=6^3$.  The bottom panel is a scatter
plot of the measured value with restored phase from complex weight
$w$, by multiplying $\frac{w}{\re w}$ for each configuration.  These
four ensembles contribute to the reweighted histogram in the upper
panel, from which we estimate the width of a single peak,
$\sigma(N_q)=2.5(3)$, at $\mu=0.139$, assuming two Gaussian peaks with
approximately equal widths.

\begin{figure}
  \centering
  \includegraphics[width=0.48\textwidth]{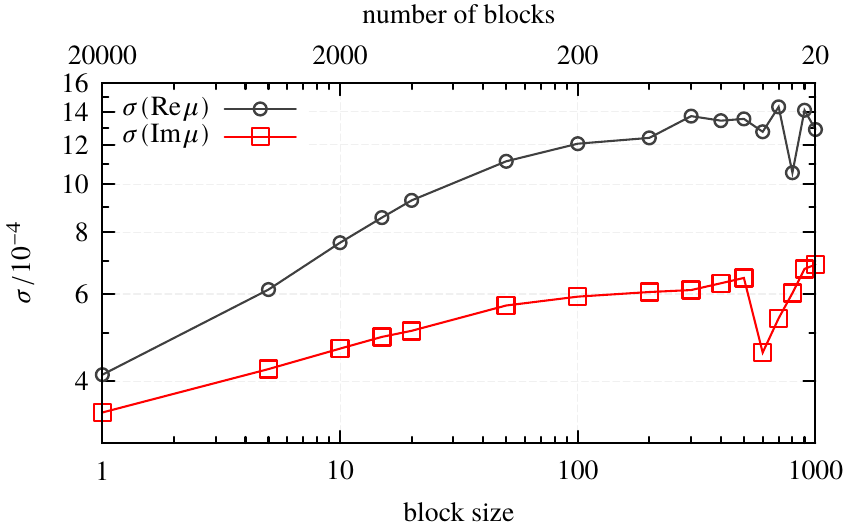}
  \hfill
  \includegraphics[width=0.48\textwidth]{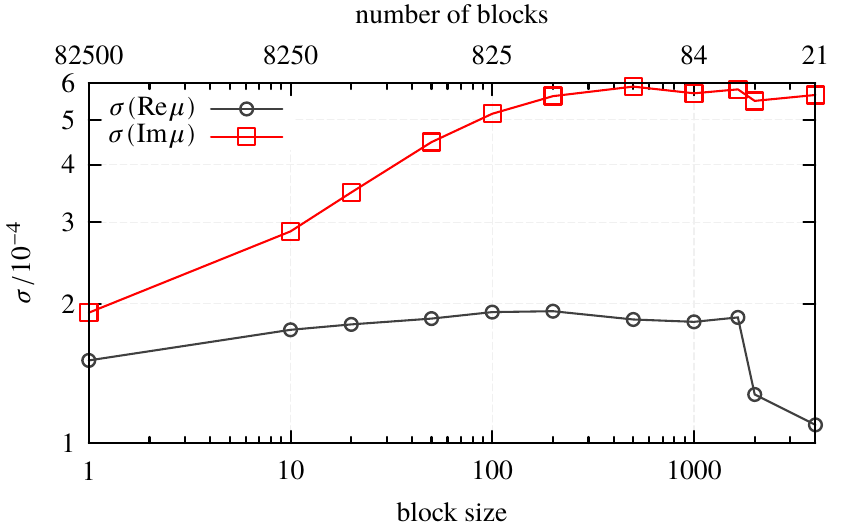}
  \caption{\label{fig:1385_zero_mu_imu_jksize}Estimated standard
    deviation against block size, for the location of the first
    partition function zero at $\bk = \bkst$, with $V=6^3$ on the left
    and $8^3$ on the right.  Both axes are in logarithmic scales.}
\end{figure}

To estimate the number of independent configurations with minimum
effect of autocorrelation, we investigate the block size effects on
the jackknife estimated standard deviation of the location of the
first partition function zero away from the real axis.  Blocking
saturates the estimated standard deviation after increasing the block
size up to about $100$, both for $20,000$ configurations for $V=6^3$
from four ensembles, and for $82,500$ configurations for $V=8^3$ from
three ensembles, shown in Figure~\ref{fig:1385_zero_mu_imu_jksize}.
The estimated error oscillates within about $10\%$, with increasing
block sizes.  Unreliable estimations seem to occur when the number of
blocks becomes $\lesssim40$; we hence choose statistical errors
estimated by jackknife with a block size of $50$.

We estimate the number of independent configurations as the total
configuration number divided by $100(20)$ (considering $20\%$ error)
following the above observation.  For the parameter set, $\bk =
\bkst$, with $V=6^3$, we have an independent sample size,
$L\sim200(40)$.  For a determination of $|\mathcal{Z}_{\text{norm}}|$
larger than three times of its standard deviation, $k=3$, applying
Eq.~\eqref{eq:1}, we get the confidence region with
$|\Delta\mu|<0.16(2)$.

\begin{figure}
  \centering
  \includegraphics[width=0.52\textwidth]{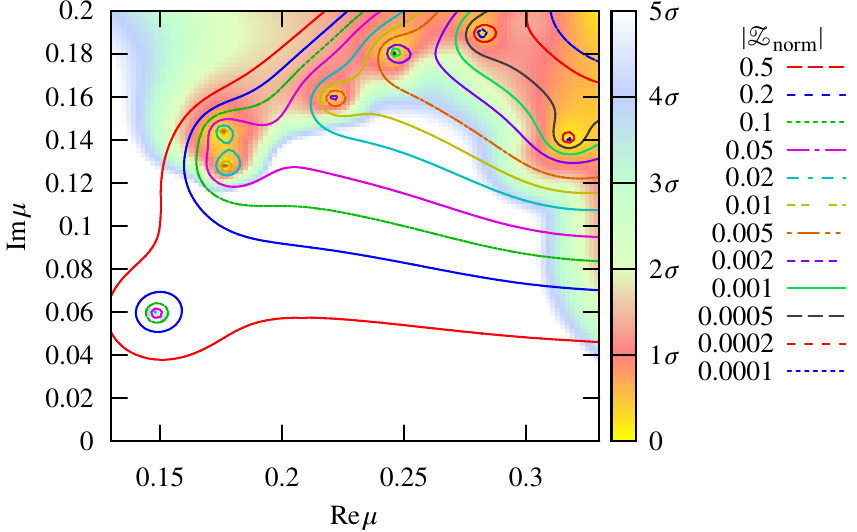}
  \caption{\label{fig:b158_nl6nt4_k01385_znorm_err}
    $|\mathcal{Z}_{\text{norm}}|$ and its standard deviation as a
    color map, at $\bk = \bkst$, with $V=6^3$.}
\end{figure}

We compare the estimated confidence radius with the statistical
uncertainty of $|\mathcal{Z}_{\text{norm}}|$, from the jackknife
method using the actual data.  As an example for $\bk = \bkst$ with
$V=6^3$, we compute $|\mathcal{Z}_{\text{norm}}|$ on a grid of the
complex $\mu$ plane, with the grid spacing $\delta\mu=0.002$.
Figure~\ref{fig:b158_nl6nt4_k01385_znorm_err} shows contours of
$|\mathcal{Z}_{\text{norm}}|$, with visible isolated zeros, which
should be expected from an analytic function.  The color map shows the
relative size of the absolute value with respect to its standard
deviation, on a scale from $0$ to $5\,\sigma$, with the pure white color
denoting $|\mathcal{Z}_{\text{norm}}|>5\,\sigma$.  In the right side of
the figure, where no real zeros of partition function from two
approximated Gaussian states would be expected, the cyan $3\,\sigma$
region is slightly larger than the edge of the confidence radius from
Gaussian approximation, $|\Delta\mu|=0.16(2)$, centering at
$\mu=0.139$.  This is probably caused by the approximation of the
partition function.  The confidence radius excludes the area where
statistical noise is too large to reliably estimate
$|\mathcal{Z}_{\text{norm}}|$, and zeros located near or outside of
the radius are mostly likely to be either spurious ones or affected by
noise.

Zeros within the confidence radius, on the other hand, are from the
cancellation of two states with high reliabilities.  The first zero
away from the real axis in
Figure~\ref{fig:b158_nl6nt4_k01385_znorm_err} is clearly a good
signal, since it is away from the boundary of our estimated confidence
region, and all values surrounding this zero is determined at more
than $5\,\sigma$ level.  The second and third one, being very close to
each other, require greater care to properly interpret their meaning.
We observe that, (a) values between these two are badly estimated
$|\mathcal{Z}_{\text{norm}}| < \sigma$, (b) some of the jackknife
blocks contain only one zero in that region, and (c) their locations
differ within statistical uncertainties estimated by the jackknife
method.  We thus consider only the lower one as the real zero and the
higher one of these two as spurious.

\begin{figure}
  \centering
  \includegraphics[width=0.52\textwidth]{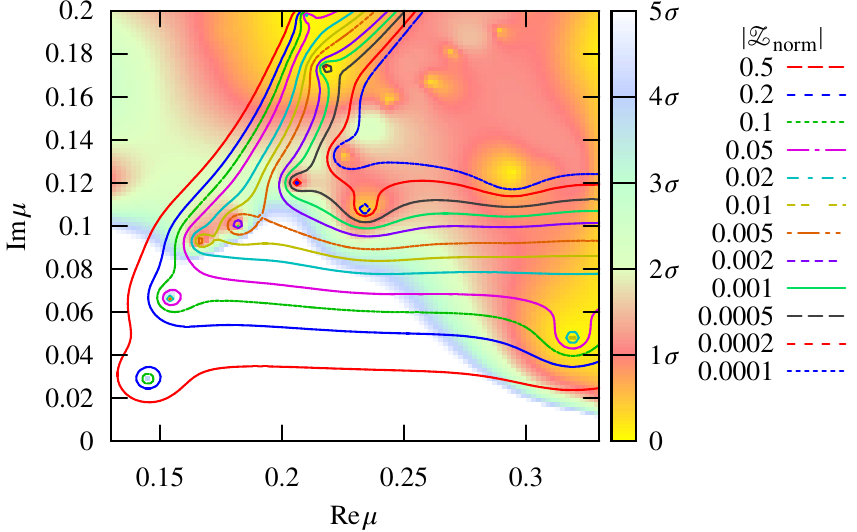}
  \caption{\label{fig:b158_nl8nt4_k01385_znorm_err}
    $|\mathcal{Z}_{\text{norm}}|$ and its standard deviation as a
    color map, at $\bk = \bkst$, with $V=8^3$.}
\end{figure}

As a further example to compare with the results from $V=6^3$, we show
the results from $V=8^3$ in
Figure~\ref{fig:b158_nl8nt4_k01385_znorm_err}.  With increasing
volume, the confidence region inferred from the color map shrinks
compared to Figure~\ref{fig:b158_nl6nt4_k01385_znorm_err}, and more
spurious-zero suspects can be seen.  Zeros, on the other hand, are
also closer to the real axis.  We can reliably determine the first two
zeros, as $|\mathcal{Z}_{\text{norm}}|$ surrounding them are nonzero
with a confidence of more than $5\,\sigma$.

We use Jackknife estimated uncertainties of zero locations in our
analysis, as they are consistent with the approximated confidence
region and offer a cross-check for the reliability of zero locations.

We also apply the above technique to zeros of partition functions
obtained with the real valued $\mu$ and the imaginary part of $\beta$.

\section{\label{sec:results}Results and discussions}

\begin{figure}
  \centering
  \includegraphics[width=0.48\textwidth]{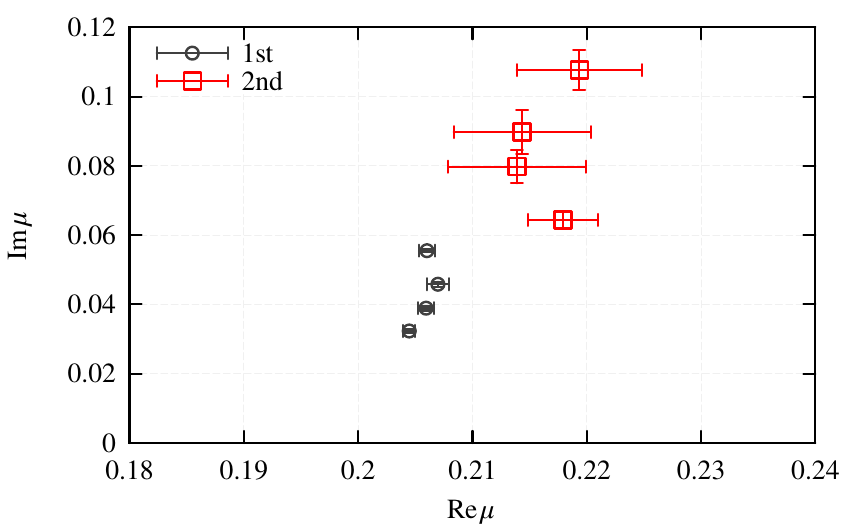}
  \hfill
  \includegraphics[width=0.48\textwidth]{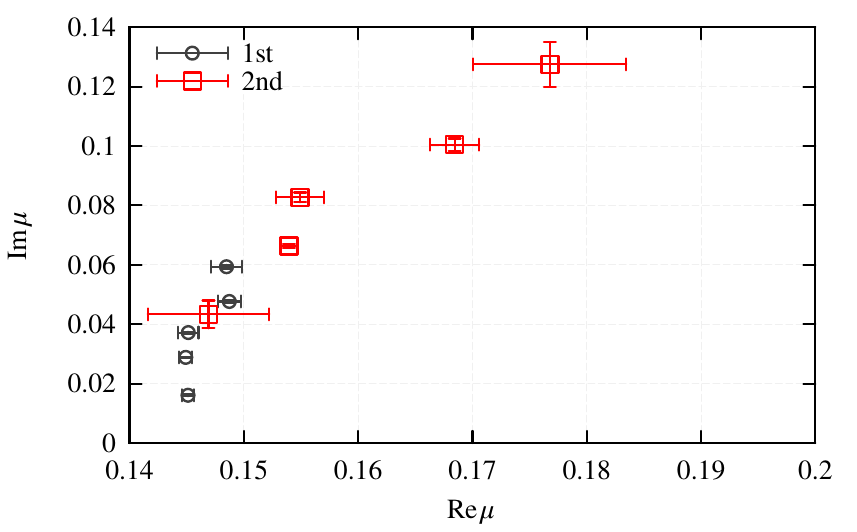}
  \caption{\label{fig:zero_mu_imu_loci}First and second zeros of
    partition functions in the plane of a complex valued $\mu$, with
    $\bk = \bkwk$ on the left and $\bk = \bkst$ on the right.  Same
    colored symbols are from $V = 6^3$, $6\cdot6\cdot8$,
    $6\cdot8\cdot8$, $8^3$, and $10^3$ (right panel only), counting
    from top to bottom.}
\end{figure}

\begin{figure}
  \centering
  \includegraphics[width=0.48\textwidth]{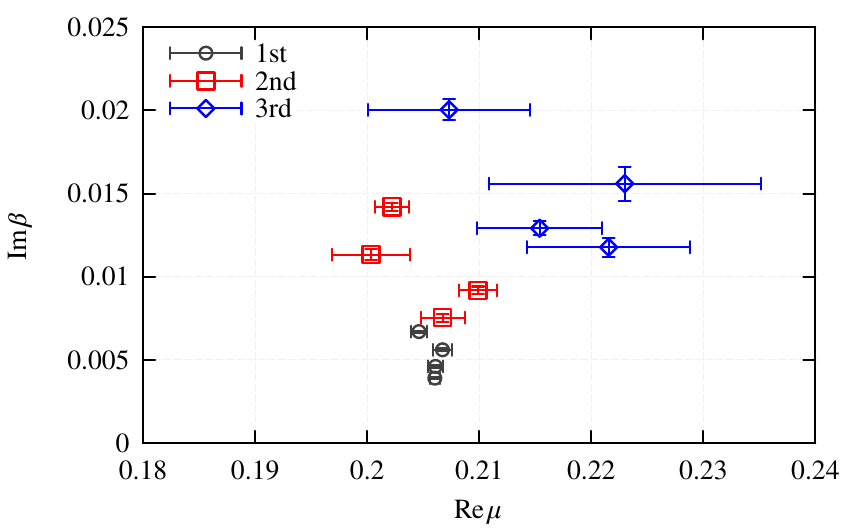}
  \hfill
  \includegraphics[width=0.48\textwidth]{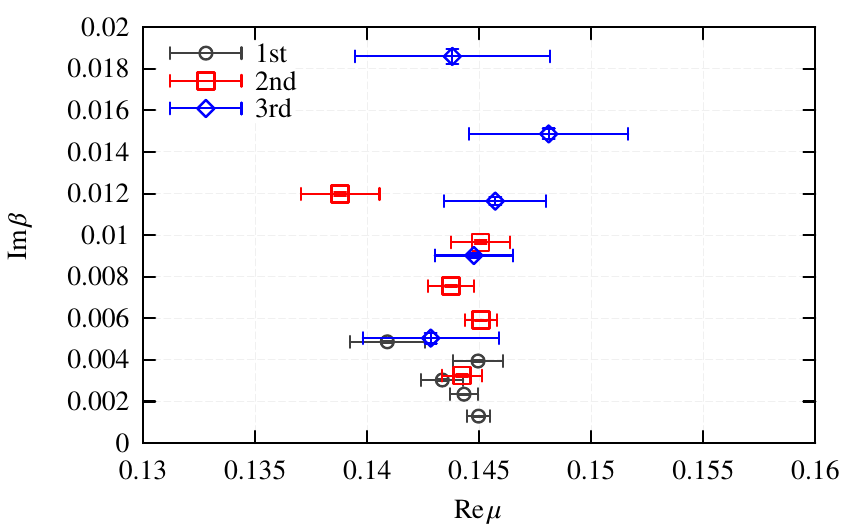}
  \caption{\label{fig:zero_mu_ibeta_loci}First three zeros of
    partition functions in the plane of a real valued $\mu$ and the
    imaginary part of $\beta$, with $(\re\beta,\,\kappa) = \bkwk$ on
    the left and $(\re\beta,\,\kappa) = \bkst$ on the right.  Same
    colored symbols are from $V = 6^3$, $6\cdot6\cdot8$,
    $6\cdot8\cdot8$, $8^3$, and $10^3$ (right panel only), counting
    from top to bottom.}
\end{figure}

We present results of partition function zeros from $\mu$ reweighting
with additional imaginary part of $\mu$ or imaginary part of $\beta$.
Figure~\ref{fig:zero_mu_imu_loci} and \ref{fig:zero_mu_ibeta_loci}
shows zero locations in the complex $\mu$ plane and in the real $\mu$
and imaginary $\beta$ plane, respectively.

Despite more ensembles with $\bk = \bkwk$, zeros with $\bk = \bkst$
possess smaller uncertainties.  This is likely due to larger phase
factors of the fermion determinant at the stronger coupling, which
goes through a transition at smaller $\mu/T$.  In addition, zeros at
the stronger coupling appear to align on a single curve, contrasting
with zeros spread out more at the weaker coupling.  This behavior may
indicate better statistics at the stronger coupling, or may suggest a
feature of zero locations for the system enters a regime of a strong
first-order transition.

We will analyze zero behaviors in a separate detailed report soon.

\acknowledgments
This work is supported in part by the Grants-in-Aid for Scientific
Research from the Ministry of Education, Culture, Sports, Science and
Technology (Nos.~23105707, 
23740177, 
22244018, 
20105002). 
The numerical calculations have been done on T2K-Tsukuba and HA-PACS
cluster system at University of Tsukuba.

\bibliographystyle{JHEP}
\bibliography{ref}

\end{document}